\documentclass[doublecol]{epl2} 

\usepackage[utf8]{inputenc}
\usepackage{hyperref,amsmath,amssymb,amsbsy,bbold,color,graphics,graphicx}
\usepackage{mathtools,relsize,mathrsfs}

\title{Rotational Heisenberg Inequalities}
\shorttitle{Rotational Heisenberg Inequalities}

\author{S. D. Brechet\inst{1} \and F. A. Reuse\inst{1} \and K. Maschke\inst{1} \and J.-P. Ansermet}
\shortauthor{S. D. Brechet \etal}

\institute{                    
  \inst{1} Institute of Condensed Matter Physics, Station 3, Ecole Polytechnique Fédérale de Lausanne - EPFL, CH-1015 Lausanne, Switzerland}

\abstract{Since their discovery in $1927$, the Heisenberg Inequalities have become an icon of quantum mechanics~\cite{Heisenberg:1927}. Often inappropriately referred to as the Uncertainty Principle, these inequalities relating the standard deviations of the position and momentum observables to Planck's constant are one of the cornerstones of the quantum formalism even if the physical interpretation of quantum mechanics remains still open to controversy nowadays~\cite{Garritz:2013}. The Heisenberg Inequalities governing translational motion are well understood. However, the corresponding inequalities pertaining to rotational motion have not been established so far. To fill this gap, we present here the Rotational Heisenberg Inequalities relating the standard deviations of the orientation axis and orbital angular momentum observables of an isolated molecule. The reason for choosing this system is that a molecule separated from its environment corresponds to a bound system preserving the orbital angular momentum.}

\begin{document}

\maketitle

\section{Relative and rest observables}

The quantum dynamics of a molecular system consisting of $N$ nuclei and $n$ electrons is obtained from the classical dynamics by applying the ``correspondence principle''. The position and momentum observables of the nucleus $\mu$ are characterised respectively by the self-adjoint operators $\boldsymbol{R}_{\mu}\,\otimes\,\mathbb{1}_{e}$ and $\boldsymbol{P}_{\mu}\,\otimes\,\mathbb{1}_{e}$, where $\mu=1,..,N$, acting trivially on the Hilbert subspace $\mathcal{H}_{e}$ associated to the electrons. Similarly, the position and momentum observables of the electron $\nu$ are respectively characterised respectively by the self-adjoint operators $\mathbb{1}_{\mathcal{N}}\,\otimes\,\boldsymbol{r}_{\nu}$ and $\mathbb{1}_{\mathcal{N}}\,\otimes\,\boldsymbol{p}_{\nu}$ where $\nu=1,..,n$, acting trivially on the Hilbert subspace $\mathcal{H}_{\mathcal{N}}$ associated to the nuclei. These operators satisfy the canonical commutation relations, i.e.
\begin{equation}\label{3.0}
\begin{split}
&\left[\ \boldsymbol{e}_{j}\cdot\boldsymbol{P}_{\mu}, \boldsymbol{e}^{k}\cdot\boldsymbol{R}_{\mu}\ \right]= -\,i\hbar\left(\boldsymbol{e}_{j}\cdot\boldsymbol{e}^{k}\right)\mathbb{1}_{\mathcal{N}}\\
&\left[\ \boldsymbol{e}_{j}\cdot\boldsymbol{p}_{\nu}, \boldsymbol{e}^{k}\cdot\boldsymbol{r}_{\nu}\ \right]= -\,i\hbar\left(\boldsymbol{e}_{j}\cdot\boldsymbol{e}^{k}\right)\mathbb{1}_{e}\,,
\end{split}
\end{equation}
where where $\boldsymbol{e}_{j}$ are the units vectors of an orthonormal basis and $\boldsymbol{e}^{k}$ are the units vectors of the dual orthonormal basis.

In order to treat molecular rotations as genuine quantum degrees of freedom, we introduce explicitly the rotation group and the associated Lie algebra~\cite{Brechet:2014}. In such a treatment, we introduce a molecular orientation operator $\boldsymbol{\omega}$ that is fully determined by the position operators $\boldsymbol{R}_{\mu}$ and $\boldsymbol{r}_{\nu}$, which ensures that it is a self-adjoint operator. Since the position operators commute, the components of the operator $\boldsymbol{\omega}$ satisfy the trivial commutation relations,
\begin{equation}\label{A.2.0}
\left[\ \boldsymbol{e}^j\cdot\boldsymbol{\omega},\ \boldsymbol{e}^k\cdot\boldsymbol{\omega}\ \right]=0\,.
\end{equation}
The orientation operator $\boldsymbol{\omega}$ should not be confused with the angle velocity operator or the phase operator. The orientation operator $\boldsymbol{\omega}$ is related to the molecular rotation operator $\mathsf{R}\left(\boldsymbol{\omega}\right)$ that belongs to the rotation group by exponentiation, i.e.
\begin{equation}\label{3.21}
\mathsf{R}\left(\boldsymbol{\omega}\right) =\exp\left(\boldsymbol{\omega}\cdot\boldsymbol{\mathsf{G}}\right)\,,
\end{equation}
taking into account the commutation relation~\eqref{A.2.0}. The components of the vector $\boldsymbol{\mathsf{G}}$ are rank-$2$ tensors and generators of the rotation group. The action of the rotation group is locally defined as,
\begin{equation}\label{3.23bis}
\left(\boldsymbol{e}_{j}\cdot\boldsymbol{\mathsf{G}}\right)\,\boldsymbol{x}=\boldsymbol{e}_{j}\times\boldsymbol{x}\,.
\end{equation}
The action of the rotation operator on a vectorial observable $\boldsymbol{A}\in\mathcal{L}\left(\mathcal{H}\right)$ is expressed in terms of the unitary representation of the rotation group $\mathsf{U}\left(\boldsymbol{\omega}\right)$ acting on the Hilbert space $\mathcal{H}$ by the well-known relation~\cite{Weinberg:2013},
\begin{equation}\label{3.24}
\mathsf{U}\left(\boldsymbol{\omega}\right)^{-1}\,\left(\boldsymbol{e}^{k}\cdot\boldsymbol{A}\right)\,\mathsf{U}\left(\boldsymbol{\omega}\right) = \left(\boldsymbol{e}^{k}\cdot\mathsf{R}\left(\boldsymbol{\omega}\right)\cdot\boldsymbol{e}_{j}\right)\,\left(\boldsymbol{e}^{j}\cdot\boldsymbol{A}\right)\,.
\end{equation}
We also introduce the operators $\boldsymbol{n}_{(j)}\left(\boldsymbol{\omega}\right)$ that are Killing vectors~\cite{Szekeres:2004} of the Lie algebra of the rotation group, which are related to the rotation operator $\mathsf{R}\left(\boldsymbol{\omega}\right)$ and the rotation generators $\boldsymbol{\mathsf{G}}$ as,
\begin{equation}\label{3.38}
\mathsf{R}\left(\boldsymbol{\omega}\right)^{-1}\cdot\left(\boldsymbol{e}_j\cdot\partial_{\boldsymbol{\omega}}\right)\,\mathsf{R}\left(\boldsymbol{\omega}\right) = \boldsymbol{n}_{(j)}\left(\boldsymbol{\omega}\right)\cdot\boldsymbol{\mathsf{G}}\,.
\end{equation}
The dual operator $\boldsymbol{m}^{(k)}\left(\boldsymbol{\omega}\right)$ satisfies the duality condition,
\begin{equation}\label{3.39}
\boldsymbol{n}_{(j)}\left(\boldsymbol{\omega}\right)\cdot\boldsymbol{m}^{(k)}\left(\boldsymbol{\omega}\right) = \boldsymbol{e}_{j}\cdot\boldsymbol{e}^{k}\,.
\end{equation}
The operators $\boldsymbol{n}_{(j)}\left(\boldsymbol{\omega}\right)$ and $\boldsymbol{m}^{(k)}\left(\boldsymbol{\omega}\right)$ determine the structure of the Lie algebra of the rotation group. 

\begin{figure}[htb]\hspace{3mm}
\begin{center}
\includegraphics[width=0.8\columnwidth]{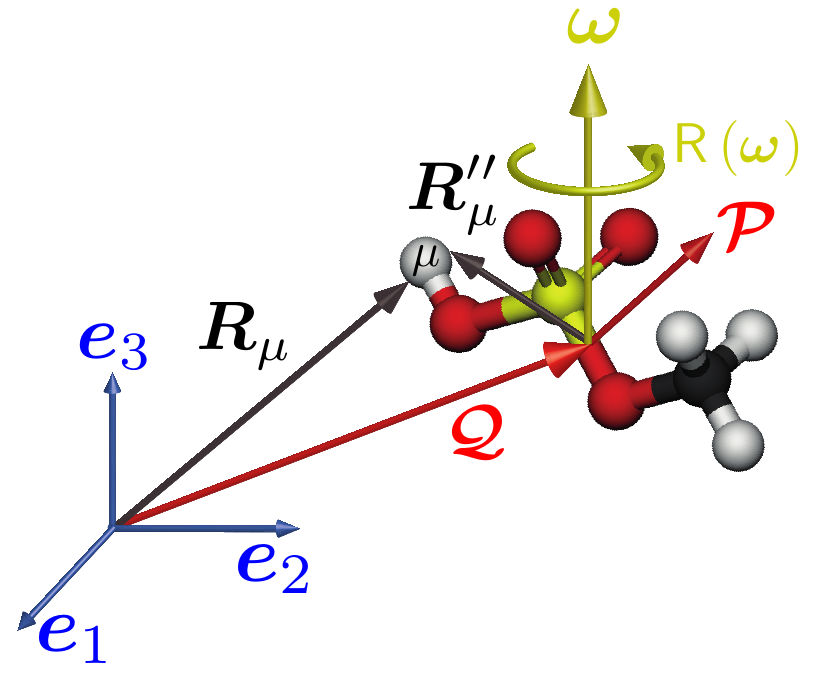}
\caption{Methyl bisulfate molecule rotating under the action of the rotation tensor $\mathsf{R}\left(\boldsymbol{\omega}\right)$ around the orientation axis $\boldsymbol{\omega}$. The vector $\boldsymbol{R}_{\mu}$ and $\boldsymbol{R}^{\prime\prime}_{\mu}$ denote respectively the position and the ``rest'' position of the nucleus $\mu$. The vectors $\boldsymbol{\mathcal{Q}}$ and $\boldsymbol{\mathcal{P}}$ denote the position and momentum of the molecular centre of mass.}
\label{frame}
\end{center}
\end{figure}

The description of molecular dynamics in a classical framework would be much simpler than in a quantum framework since in the former a rest frame could be attached easily to the physical system. In quantum physics, the approach is slightly different because observables are described mathematically by operators, which implies that there exists no rest frame and no centre of mass frame associated to the molecular system. However, even in the absence of a centre of mass frame, the position and momentum observables of the centre of mass can be expressed mathematically as self-adjoint operators. This enables us to define other position and momentum observables with respect to the centre of mass. We shall refer to them as ``relative'' position and momentum observables because they are the quantum equivalent of the classical relative position and momentum variables defined with respect to the center of mass frame. Then, using the rotation operator, we define the ``rest'' position and momentum observables, which are the quantum equivalent of the classical position and momentum variables defined in the molecular rest frame. The description of molecular dynamics is illustrated in Fig~\ref{frame}.

Applying the correspondence principle, the position, momentum and angular momentum observables associated to the center of mass are respectively given by the self-adjoint operators, 
\begin{equation}\label{3.3}
\begin{split}
&\boldsymbol{\mathcal{Q}}=\frac{1}{\mathcal{M}}\left(\ \sum_{\mu=1}^{N} M_{\mu}\,\boldsymbol{R}_{\mu}\otimes\mathbb{1}_{e}+\sum_{\nu=1}^{n}\,\mathbb{1}_{\mathcal{N}}\otimes m\,\boldsymbol{r}_{\nu}\ \right)\\
&\boldsymbol{\mathcal{P}}=\sum_{\mu=1}^{N}\boldsymbol{P}_{\mu}\otimes\mathbb{1}_{e}+\sum_{\nu=1}^{n}\,\mathbb{1}_{\mathcal{N}}\otimes\boldsymbol{p}_{\nu}\,,
\end{split}
\end{equation}
where $M_{\mu}$ is the mass of the nucleus $\mu$, $m$ is the mass of the electron, $\mathcal{M}= M + nm$ is the mass of the molecule defined in terms of the mass $M$ of the nuclei. The definitions~\eqref{3.3} and the commutation relations~\eqref{3.0} imply that the operators $\boldsymbol{\mathcal{P}}$ and $\boldsymbol{\mathcal{Q}}$ satisfy the commutation relation,
\begin{equation}
\label{3.6}
\left[\ \boldsymbol{e}_{j}\cdot\boldsymbol{\mathcal{P}}, \boldsymbol{e}^{k}\cdot\boldsymbol{\mathcal{Q}}\ \right]= -\,i\hbar\left(\boldsymbol{e}_{j}\cdot\boldsymbol{e}^{k}\right)\mathbb{1}\,.
\end{equation}

The ``relative'' position operators $\boldsymbol{R}^{\prime}_{\mu}$ and $\boldsymbol{r}^{\prime}_{\nu}$, and the ``relative'' momentum operators $\boldsymbol{P}^{\prime}_{\mu}$ and $\boldsymbol{p}^{\prime}_{\nu}$ are related respectively to the operators $\boldsymbol{R}_{\mu}$, $\boldsymbol{r}_{\nu}$, $\boldsymbol{P}_{\mu}$ and $\boldsymbol{p}_{\nu}$ by
\begin{equation}\label{3.9}
\begin{split}
&\boldsymbol{R}^{\prime}_{\mu}=\boldsymbol{R}_{\mu}\otimes\mathbb{1}_{e}-\boldsymbol{\mathcal{Q}}\,,\vphantom{\frac{m}{\mathcal{M}}}\\
&\boldsymbol{r}^{\prime}_{\nu}=\mathbb{1}_{\mathcal{N}}\otimes\boldsymbol{r}_{\nu}-\boldsymbol{\mathcal{Q}}\,,\vphantom{\frac{m}{\mathcal{M}}}\\
&\boldsymbol{P}^{\prime}_{\mu}=\boldsymbol{P}_{\mu}\otimes\mathbb{1}_{e}-\frac{M_{\mu}}{\mathcal{M}}\ \boldsymbol{\mathcal{P}}\,,\\
&\boldsymbol{p}^{\prime}_{\nu}=\mathbb{1}_{\mathcal{N}}\otimes\boldsymbol{p}_{\nu}-\frac{m}{\mathcal{M}}\ \boldsymbol{\mathcal{P}}\,.
\end{split}
\end{equation}
and satisfy the conditions,
\begin{equation}\label{3.11}
\begin{split}
&\sum_{\mu=1}^{N}\,M_{\mu}\,\boldsymbol{R}^{\prime}_{\mu}+\sum_{\nu=1}^{n} m\ \boldsymbol{r}^{\prime}_{\nu} = \boldsymbol{0}\,,\\
&\sum_{\mu=1}^{N}\,\boldsymbol{P}^{\prime}_{\mu}+\sum_{\nu=1}^{n}\,\boldsymbol{p}^{\prime}_{\nu} = \boldsymbol{0}\,,
\end{split}
\end{equation}
that are a direct consequence of the definitions~\eqref{3.3} and~\eqref{3.9}. The definitions~\eqref{3.9} and the canonical commutations relations~\eqref{3.0} and~\eqref{3.6} imply that the canonical commutations relations between the ``relative'' observables are given by,  
\begin{equation}\label{3.13}
\begin{split}
&\left[\ \boldsymbol{e}_j\cdot\boldsymbol{P}^{\prime}_{\mu},\  \boldsymbol{e}^k\cdot\boldsymbol{R}^{\prime}_{\nu}\ \right] =  -\,i\hbar\left(\boldsymbol{e}_j\cdot\boldsymbol{e}^k\right)\left( \delta_{\mu\nu}-\,\frac{M_{\mu}}{\mathcal{M}} \right)\,\mathbb{1}\,,\\
&\left[\ \boldsymbol{e}_j\cdot\boldsymbol{p}^{\prime}_{\mu},\  \boldsymbol{e}^k\cdot\boldsymbol{r}^{\prime}_{\nu}\ \right] = -\,i\hbar\left(\boldsymbol{e}_j\cdot\boldsymbol{e}^k\right)\left( \delta_{\mu\nu}-\,\frac{m}{\mathcal{M}} \right)\,\mathbb{1}\,.
\end{split}
\end{equation}

The ``rest'' position operators $\boldsymbol{R}^{\prime\prime}_{\mu}$ and $\boldsymbol{r}^{\prime\prime}_{\nu}$, and the ``rest'' momentum operators $\boldsymbol{P}^{\prime\prime}_{\mu}$ and $\boldsymbol{p}^{\prime\prime}_{\nu}$ are related respectively to the operators ``relative'' $\boldsymbol{R}^{\prime}_{\mu}$, $\boldsymbol{r}^{\prime}_{\nu}$, $\boldsymbol{P}^{\prime}_{\mu}$ and $\boldsymbol{p}^{\prime}_{\nu}$ by 
\begin{equation}\label{3.13bis}
\begin{split}
&\boldsymbol{e}^{j}\cdot\boldsymbol{R}^{\prime\prime}_{\mu} = \left(\boldsymbol{e}^{j}\cdot\mathsf{R}\left(\boldsymbol{\omega}\right)^{-1}\cdot\boldsymbol{e}_{k}\right)\,\left(\boldsymbol{e}^{k}\cdot\boldsymbol{R}^{\prime}_{\mu}\right)\,,\\
&\boldsymbol{e}^{j}\cdot\boldsymbol{r}^{\prime\prime}_{\nu} =  \left(\boldsymbol{e}^{j}\cdot\mathsf{R}\left(\boldsymbol{\omega}\right)^{-1}\cdot\boldsymbol{e}_{k}\right)\,\left(\boldsymbol{e}^{k}\cdot\boldsymbol{r}^{\prime}_{\nu}\right)\,,\\
&\boldsymbol{e}_{j}\cdot\boldsymbol{P}^{\prime\prime}_{\mu} = \frac{1}{2}\,\left\{\ \boldsymbol{e}^{k}\cdot\mathsf{R}\left(\boldsymbol{\omega}\right)\cdot\boldsymbol{e}_{j},\ \boldsymbol{e}_k\cdot\boldsymbol{P}^{\prime}_{\mu}\ \right\}\,,\\
&\boldsymbol{e}_{j}\cdot\boldsymbol{p}^{\prime\prime}_{\nu} = \left(\boldsymbol{e}^{k}\cdot\mathsf{R}\left(\boldsymbol{\omega}\right)\cdot\boldsymbol{e}_{j}\right)\,\left(\boldsymbol{e}_{k}\cdot\boldsymbol{p}^{\prime}_{\nu}\right)\,.\vphantom{\frac{1}{2}}
\end{split}
\end{equation}
where $\mathsf{R}\left(\boldsymbol{\omega}\right)$ is a rotation operator, which is a tensorial operator that is a function of the pseudo-vectorial operator $\boldsymbol{\omega}$ describing the orientation of the molecular system. The brackets $\{\ ,\ \}$ in the definitions~\eqref{3.13bis} denote an anticommutator accounting for the fact that the rotation operator $\mathsf{R}\left(\boldsymbol{\omega}\right)$ does not commute with the position operator $\boldsymbol{P}^{\prime}_{\mu}$ of the nuclei. In the definitions~\eqref{3.13bis}, we used the Einstein summation convention on alternated indices. The ``rest'' observables satisfy the conditions,
\begin{equation}\label{3.11bis}
\begin{split}
&\sum_{\mu=1}^{N}\,M_{\mu}\,\boldsymbol{R}^{\prime\prime}_{\mu} + \sum_{\nu=1}^{n}\,m\,\boldsymbol{r}^{\prime\prime}_{\nu} = \boldsymbol{0}\,,\\
&\sum_{\mu=1}^{N}\,\boldsymbol{P}^{\prime\prime}_{\mu}+\sum_{\nu=1}^{n}\,\boldsymbol{p}^{\prime\prime}_{\nu} = \boldsymbol{0}\,,
\end{split}
\end{equation}
that are a direct consequence of the conditions~\eqref{3.11} and the definitions~\eqref{3.13bis}. 

The definitions~\eqref{3.13bis}, the commutation relations~\eqref{3.13}, the rotational group action~\eqref{3.23bis} and the relation~\eqref{3.38} imply that the canonical commutation relations between the ``rest'' observables are given by,~\cite{Brechet:2014}
\begin{align}\label{3.17quad}
&\left[\ \boldsymbol{e}_j\cdot\boldsymbol{P}^{\prime\prime}_{\mu},\  \boldsymbol{e}^k\cdot\boldsymbol{R}^{\prime\prime}_{\nu}\ \right] = -\,i\hbar\left(\boldsymbol{e}_j\cdot\boldsymbol{e}^k\right)\left(\delta_{\mu\nu}-\frac{M_{\mu}}{\mathcal{M}}\right)\,\mathbb{1}\nonumber\\ &-\left[\,\boldsymbol{e}_{j}\cdot\boldsymbol{P}^{\prime\prime}_{\mu},\,\boldsymbol{e}^{\ell}\cdot\boldsymbol{\omega}\,\right]\,\boldsymbol{e}^{k}\cdot\left(\boldsymbol{n}_{(\ell)}\left(\boldsymbol{\omega}\right)\times\boldsymbol{R}^{\prime\prime}_{\nu}\right)\,,\\
&\left[\ \boldsymbol{e}_j\cdot\boldsymbol{p}^{\prime\prime}_{\mu},\  \boldsymbol{e}^k\cdot\boldsymbol{r}^{\prime\prime}_{\nu}\ \right] = -\,i\hbar\left(\boldsymbol{e}_j\cdot\boldsymbol{e}^k\right)\left(\delta_{\mu\nu}-\frac{m}{\mathcal{M}}\right)\,\mathbb{1}\,,
\end{align}
where the structural differences are due to the fact that the nuclear ``rest'' momentum $\boldsymbol{P}^{\prime\prime}_{\mu}$ does not commute with the molecular orientation operator  $\boldsymbol{\omega}$ whereas the electronic ``rest'' momentum $\boldsymbol{p}^{\prime\prime}_{\nu}$ does.

\section{Internal observables}

The ``rest'' position and momentum observables can be recast in terms of internal observables characterising the vibrational, rotational and electronic degrees of freedom of the quantum molecular system. In order to do so, we introduce the scalar operators $Q^{\alpha}$, where $\alpha = 1,..,3N-6$, characterising the deformation amplitude of the vibrational modes of the $N$ nuclei and the vectorial operators $\boldsymbol{q}_{(\nu)}$, where $\nu = 1,..,n$, related to the position of the electrons. The ``rest'' position operators $\boldsymbol{R}^{\prime\prime}_{\mu}$ and $\boldsymbol{r}^{\prime\prime}_{\mu}$ are expressed in terms of the operators $Q^{\alpha}$, $\boldsymbol{q}_{(\nu)}$ and the equilibrium configuration $\boldsymbol{R}^{(0)}_{\mu}$ of the nucleus $\mu$, i.e.~\cite{Brechet:2014}
\begin{align}\label{3.14}
&\boldsymbol{R}^{\prime\prime}_{\mu} = \boldsymbol{R}^{(0)}_{\mu}\,\mathbb{1}+\frac{1}{\sqrt{M_{\mu}}}\,Q^{\alpha}\,\boldsymbol{X}_{\mu\alpha} -\,\frac{m}{M}\!\sum_{\nu,\nu^{\prime} = 1}^{n}\!A_{\nu\nu^{\prime}}\,\boldsymbol{q}_{(\nu^{\prime})}\,,\nonumber\\
&\boldsymbol{r}^{\prime\prime}_{\nu} = \sum_{\nu^{\prime} = 1}^{n}\,A_{\nu\nu^{\prime}}\,\boldsymbol{q}_{(\nu^{\prime})}\,,
\end{align}
where we used Einstein's implicit summation convention for the vibrational modes $\alpha$, and the matrix elements $A_{\nu\nu^{\prime}}$ are defined as,
\begin{equation}\label{3.15.A}
A_{\nu\nu^{\prime}} \equiv \delta_{\nu\nu^{\prime}} + \frac{1}{n}\,\left(\sqrt{\frac{M}{\mathcal{M}}}-\,1\right)\,.
\end{equation}
Similarly, the ``rest'' momentum operators $\boldsymbol{P}^{\prime\prime}_{\mu}$ and are expressed in terms of the operators $P_{\alpha}$, $\boldsymbol{p}_{(\nu^{\prime})}$ and the angular velocity pseudo-vectorial operator $\boldsymbol{\Omega}$, i.e.~\cite{Brechet:2014}
\begin{align}\label{3.25}
&\boldsymbol{P}^{\prime\prime}_{\mu} = \boldsymbol{\Omega}\times\left(M_{\mu}\,\boldsymbol{R}^{(0)}_{\mu}\right) + \sqrt{M_{\mu}}\,P_{\alpha}\,\boldsymbol{X}^{\alpha}_{\mu}\nonumber\\
&\phantom{\boldsymbol{P}^{\prime\prime}_{\mu} =}-\,\frac{M_{\mu}}{M}\!\sum_{\nu,\nu^{\prime} = 1}^{n}\!A_{\nu\nu^{\prime}}\,\boldsymbol{p}_{(\nu^{\prime})}\,,\nonumber\\
&\boldsymbol{p}^{\prime\prime}_{\nu} = \sum_{\nu^{\prime} = 1}^{n}\,A_{\nu\nu^{\prime}}\,\boldsymbol{p}_{(\nu^{\prime})}\,.
\end{align}
The definition~\eqref{3.15.A} is not unique and the particular choice made here leads to the usual structure~\eqref{3.32.1} of the canonical commutation relation between the operators $\boldsymbol{q}_{(\nu)}$ and $\boldsymbol{p}_{(\nu)}$. The vector set $\{\boldsymbol{X}_{\mu\alpha}\}$ is the basis characterising the vibrational modes that is orthonormal to the dual orthonormal basis $\{\boldsymbol{X}^{\beta}_{\mu}\}$, i.e.
\begin{equation}\label{3.18}
\sum_{\mu=1}^{N}\,\boldsymbol{X}_{\mu\alpha}\cdot\boldsymbol{X}^{\beta}_{\mu} = \delta^{\beta}_{\alpha}\,.
\end{equation}
In order for the identities~\eqref{3.14} and~\eqref{3.25} to satisfy the conditions~\eqref{3.11bis}, we need to impose conditions on the vectors $\boldsymbol{R}^{(0)}_{\mu}$ and $\boldsymbol{X}_{\mu\alpha}$. First, we choose the origin of the coordinate system such that it coincides with the center of mass at equilibrium, i.e.
\begin{equation}\label{3.16}
\sum_{\mu=1}^{N} M_{\mu}\,\boldsymbol{R}^{(0)}_{\mu} = \boldsymbol{0}\,.\\
\end{equation}
Then, we require the deformation modes of the molecule to preserve the momentum, i.e.
\begin{equation}\label{3.17}
\sum_{\mu=1}^{N}\,\sqrt{M_{\mu}}\,\boldsymbol{X}_{\mu\alpha}=\boldsymbol{0}\,.
\end{equation}
We also require the deformation modes of the molecule to preserve the orbital angular momentum, i.e.
\begin{equation}\label{3.17bis}
\sum_{\mu=1}^{N} \sqrt{M_{\mu}}\,\left(\boldsymbol{R}^{(0)}_{\mu}\times\boldsymbol{X}_{\mu\alpha}\right)=\boldsymbol{0}\,.
\end{equation}
The constraints~\eqref{3.16}-\eqref{3.17bis} are known as the Eckart conditions~\cite{Eckart:1935}. Finally, we choose the orientation of the coordinate system such that the inertia tensor of the equilibrium position of the nuclei is diagonal, i.e.
\begin{equation}\label{3.16bis}
\begin{split}
&\sum_{\mu=1}^{N} M_{\mu}\left(\boldsymbol{e}_j\cdot\boldsymbol{R}^{(0)}_{\mu}\right)\left(\boldsymbol{e}_k\cdot\boldsymbol{R}^{(0)}_{\mu}\right)\\
&= \sum_{\mu=1}^{N} M_{\mu}\left(\boldsymbol{e}_j\cdot\boldsymbol{e}_k\right)\left(\boldsymbol{e}_k\cdot\boldsymbol{R}^{(0)}_{\mu}\right)^2\,.
\end{split}
\end{equation}
The first relation~\eqref{3.13bis} and the physical constraints~\eqref{3.16} and~\eqref{3.17bis} determine the rotation operator $\mathsf{R}\left(\boldsymbol{\omega}\right)$, i.e.
\begin{equation}\label{3.39 ter}
\sum_{\mu=1}^{N}\, M_{\mu}\,\boldsymbol{R}^{(0)}_{\mu}\times\left(\mathsf{R}\left(\boldsymbol{\omega}\right)^{-1}\cdot\boldsymbol{R}^{\prime}_{\mu}\right)=\boldsymbol{0}\,.
\end{equation}
\begin{figure}[htb]\hspace{3mm}
\begin{center}
\includegraphics[width=0.80\columnwidth]{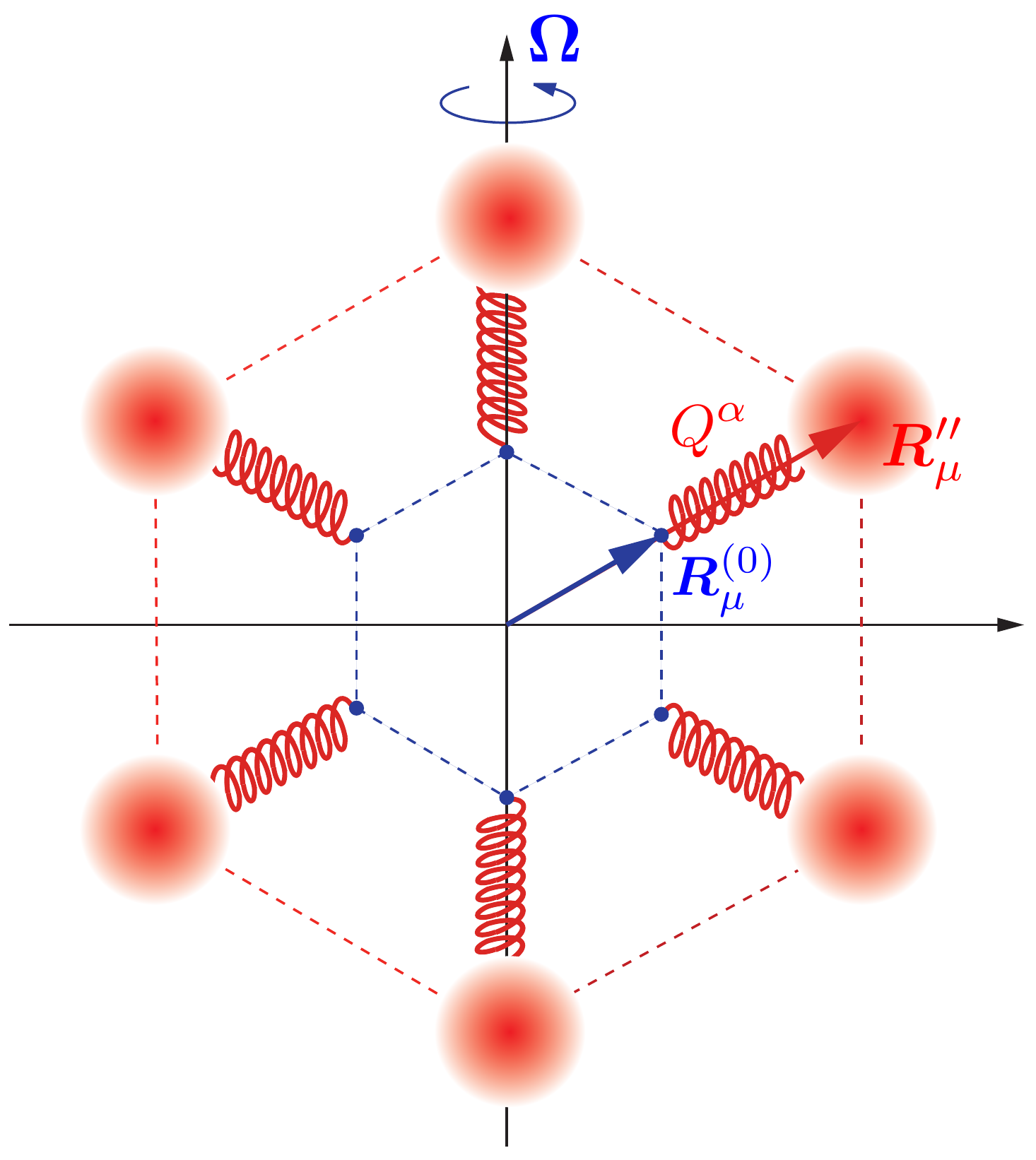}
\caption{Molecule rotating with an angular velocity $\boldsymbol{\Omega}$ around the vertical axis coinciding with the orientation $\boldsymbol{\omega}$. The vector $\boldsymbol{R}^{(0)}_{\mu}$ and the vectorial operator $\boldsymbol{R}^{\prime\prime}_{\mu}$ are respectively the equilibrium configuration and ``rest'' position of the nucleus $\mu$. The scalar $Q_{\alpha}$ is the molecular vibration mode $\alpha$.}
\label{molecule}
\end{center}
\end{figure}

To emphasize the physical motivation behind the previous formal development, we consider the classical counterpart of a quantum molecular system illustrated on Fig~\ref{molecule}. In a classical framework, the classical counterpart of the operatorial relation~\eqref{3.39 ter} determines the rest frame of the molecular system. Moreover, the equilibrium configuration of a molecule is given by a vector set $\{\boldsymbol{R}^{(0)}_{\mu}\}$ describing the position of the nuclei. The condition~\eqref{3.16} implies that the centre of mass of the molecule coincides with the origin of the coordinate system and the condition~\eqref{3.16bis} requires the inertial tensor of this molecule to be diagonal with respect to the coordinate system. The set of orthonormal vectors $\{\boldsymbol{X}_{\mu\alpha}\}$ characterise the $3N-6$ normal deformation modes of the molecule and thus account for the vibrations. The condition~\eqref{3.17} implies that the normal deformation modes preserve the momentum of the molecule and the condition~\eqref{3.17bis} requires that these modes also to preserve the orbital angular momentum.

The internal observables are described by the scalar operators $Q^{\alpha}$, $P_{\alpha}$, the vectorial operators $\boldsymbol{q}_{(\nu)}$, $\boldsymbol{p}_{(\nu)}$ and the pseudo-vectorial operators $\boldsymbol{\Omega}$ and $\boldsymbol{\omega}$. The inversion of the definitions~\eqref{3.14} and~\eqref{3.25} yields explicit expressions for the internal observables $Q^{\alpha}$, $P_{\alpha}$, $\boldsymbol{q}_{(\nu)}$ and $\boldsymbol{p}_{(\nu)}$, i.e.
\begin{equation}\label{3.27 prime}
\begin{split}
&Q^{\alpha} = \sum_{\mu=1}^{N}\,\sqrt{M_{\mu}}\, \boldsymbol{X}^{\alpha}_{\mu}\cdot\Big(\boldsymbol{R}^{\prime\prime}_{\mu}-\,\boldsymbol{R}^{(0)}_{\mu}\,\mathbb{1}\Big)\,,\\
&\boldsymbol{q}_{(\nu)} = \sum_{\nu^{\prime}=1}^{n}\left(\delta_{\nu\nu^{\prime}} + \frac{1}{n}\,\left(\sqrt{\frac{\mathcal{M}}{M}}-\,1\right)\right)\,\boldsymbol{r}^{\prime\prime}_{\nu^{\prime}}\,,\\
&P_{\alpha} = \sum_{\mu=1}^{N}\frac{1}{\sqrt{M_{\mu}}}\,\left(\boldsymbol{X}_{\mu\alpha}\cdot\boldsymbol{P}^{\prime\prime}_{\mu}\right)\,,\\
&\boldsymbol{p}_{(\nu)} = \sum_{\nu^{\prime}=1}^{n}\left(\delta_{\nu\nu^{\prime}} + \frac{1}{n}\,\left(\sqrt{\frac{\mathcal{M}}{M}}-\,1\right)\right)\,\boldsymbol{p}^{\prime\prime}_{\nu^{\prime}}\,.
\end{split}
\end{equation}
The expressions~\eqref{3.27 prime}, the commutations relations~\eqref{3.17quad} between the ``rest'' observables and the Eckart conditions~\eqref{3.16}-\eqref{3.17bis} yield the canonical commutation relations relations between the vibrational internal observables and the electronic internal observables respectively, i.e.
\begin{equation}\label{3.32.1}
\begin{split}
&\left[\ P_{\alpha},\ Q^{\beta}\ \right] = -\,i\hbar\,\delta^{\beta}_{\alpha}\,\mathbb{1}\,,\\
&\left[\ \boldsymbol{e}_{j}\cdot\boldsymbol{p}_{(\nu)},\ \boldsymbol{e}^{k}\cdot\boldsymbol{q}_{(\nu^{\prime})}\ \right] = -\,i\hbar\,\left(\boldsymbol{e}_{j}\cdot\boldsymbol{e}^{k}\right)\,\delta_{\nu\nu^{\prime}}\,\mathbb{1}\,.
\end{split}
\end{equation}

\section{Orbital angular momentum observable}

The ``relative'' orbital angular momentum operator $\boldsymbol{L}^{\prime}$ is defined as,
\begin{equation}\label{3.17pet}
\boldsymbol{L}^{\prime} = \sum_{\mu=1}^{N}\,\boldsymbol{R}^{\prime}_{\mu}\times\boldsymbol{P}^{\prime}_{\mu} + \sum_{\nu=1}^{n}\,\boldsymbol{r}^{\prime}_{\nu}\times\boldsymbol{p}^{\prime}_{\nu}\,,
\end{equation}
and the ``rest'' orbital angular momentum operator $\boldsymbol{L}$ is defined as,
\begin{equation}\label{3.26bis}
\boldsymbol{L} = \frac{1}{2}\,\sum_{\mu=1}^{N}\,\left[\ \boldsymbol{R}^{\prime\prime}_{\mu},\ \boldsymbol{P}^{\prime\prime}_{\mu}\ \right]_{\boldsymbol{\times}} + \frac{1}{2}\,\sum_{\nu=1}^{n}\,\left[\ \boldsymbol{r}^{\prime\prime}_{\nu},\ \boldsymbol{p}^{\prime\prime}_{\nu}\ \right]_{\boldsymbol{\times}}\,,
\end{equation}
where we used the notation convention $\left[\ \boldsymbol{A},\ \boldsymbol{B}\ \right]_{\boldsymbol{\times}} = \boldsymbol{A}\times\boldsymbol{B} -\,\boldsymbol{B}\times\boldsymbol{A}$. 

In order to express ``rest'' orbital angular momentum operator $\boldsymbol{L}$ in terms of the internal observables, we introduce the the inertia tensorial operator $\mathsf{I}\left(Q^{\,\boldsymbol{.}}\right)$. The components of the operator are defined as,
\begin{equation}\label{3.31}
\boldsymbol{e}_{k}\cdot\mathsf{I}\left(Q^{\,\boldsymbol{.}}\right)\cdot\boldsymbol{e}_{\ell} = \left(\boldsymbol{e}_{k}\cdot\mathsf{I}_{0}\cdot\boldsymbol{e}_{\ell}\right)\,\mathbb{1} + Q^{\alpha}\,\left(\boldsymbol{e}_{k}\cdot\mathsf{I}_{\alpha}\cdot\boldsymbol{e}_{\ell}\right)\,,
\end{equation}
where the dot in the argument of the operator $\mathsf{I}\left(Q^{\,\boldsymbol{.}}\right)$ refers to all the vibrational modes. The first term on the RHS of the definition~\eqref{3.31} is required to be diagonal with respect to the rotating molecular system according to the constraint~\eqref{3.16bis}, i.e.
\begin{align}\label{3.32bis}
&\boldsymbol{e}_{k}\cdot\mathsf{I}_{0}\cdot\boldsymbol{e}_{\ell} = \left(\boldsymbol{e}_{k}\cdot\mathsf{I}_{0}\cdot\boldsymbol{e}_{k}\right)\left(\boldsymbol{e}_{k}\cdot\boldsymbol{e}_{\ell}\right)\\
&\phantom{\boldsymbol{e}_{k}\cdot\mathsf{I}_{0}\cdot\boldsymbol{e}_{\ell}}= \sum_{\mu=1}^{N}\,M_{\mu}\,\left({\boldsymbol{R}^{(0)}_{\mu}}^{2}-\,\left(\boldsymbol{e}_{k}\cdot\boldsymbol{R}^{(0)}_{\mu}\right)^{2}\right)\left(\boldsymbol{e}_{k}\cdot\boldsymbol{e}_{\ell}\right)\,,\nonumber
\end{align}
and the second term on the RHS, i.e.
\begin{align}\label{3.33}
&\boldsymbol{e}_{k}\cdot\mathsf{I}_{\alpha}\cdot\boldsymbol{e}_{\ell} = \sum_{\mu=1}^{N}\,\sqrt{M_{\mu}}\, \left(\boldsymbol{e}_{k}\times\boldsymbol{X}_{\mu\alpha}\right)\cdot\left(\boldsymbol{e}_{\ell}\times\boldsymbol{R}^{(0)}_{\mu}\right)\nonumber\\
&\phantom{\boldsymbol{e}_{k}\cdot\mathsf{I}_{\alpha}\cdot\boldsymbol{e}_{\ell}} = \boldsymbol{e}_{\ell}\cdot\mathsf{I}_{\alpha}\cdot\boldsymbol{e}_{k}\,,
\end{align}
is symmetric under the condition~\eqref{3.17bis}. Using the definitions~\eqref{3.14}-\eqref{3.25} and~\eqref{3.31}-\eqref{3.33} the ``rest'' orbital angular momentum~\eqref{3.26bis} is recast as,~\cite{Brechet:2014}
\begin{align}\label{3.27}
&\boldsymbol{L} = \frac{1}{2}\,\Big\{\ \mathsf{I}\left(Q^{\,\boldsymbol{.}}\right),\ \boldsymbol{\Omega}\ \Big\}_{\mathsmaller{\bullet}} + \frac{1}{2}\,\sum_{\mu = 1}^{N}\,\left[\ Q^{\alpha}\,\boldsymbol{X}_{\mu\alpha},\ P_{\beta}\,\boldsymbol{X}^{\beta}_{\mu}\ \right]_{\times}\nonumber\\
&\phantom{\boldsymbol{L} =} + \frac{1}{2}\,\sum_{\nu = 1}^{n}\,\left[\ \boldsymbol{q}_{(\nu)},\ \boldsymbol{p}_{(\nu)}\ \right]_{\boldsymbol{\times}}\,,
\end{align}
where we used the convention $\left\{\ \boldsymbol{A},\ \boldsymbol{B}\ \right\}_{\mathsmaller{\bullet}} = \boldsymbol{A}\cdot\boldsymbol{B} + \boldsymbol{B}\cdot\boldsymbol{A}$. In expression~\eqref{3.27}, the first term on the RHS represents the molecular orbital angular momentum, the second term corresponds to the deformation orbital angular momentum and the last term is the electronic orbital angular momentum.

The orbital angular momentum operator $\boldsymbol{L}$ commutes with the operators $Q^{\alpha}$, $P_{\beta}$, $\boldsymbol{q}_{(\nu)}$ and $\boldsymbol{p}_{(\nu)}$ but it does not commute with the orientation operator $\boldsymbol{\omega}$. The commutation relation between the angular velocity operator $\boldsymbol{\Omega}$ and the molecular orientation operator $\boldsymbol{\omega}$ is given by,~\cite{Brechet:2014}
\begin{equation}\label{3.33.1}
\left[\ \boldsymbol{e}^{j}\cdot\boldsymbol{\Omega},\ \boldsymbol{e}^k\cdot\boldsymbol{\omega}\ \right] = -\,i\hbar\,\left(\boldsymbol{e}^{j}\cdot\mathsf{I}\left(Q^{\,\boldsymbol{.}}\right)^{-1}\cdot\boldsymbol{m}^{(k)}\left(\boldsymbol{\omega}\right)\right)\,.
\end{equation}
The definition~\eqref{3.27} and the commutation relation~\eqref{3.33.1} yield the commutation relation between $\boldsymbol{L}$ and $\boldsymbol{\omega}$, i.e.
\begin{equation}\label{3.35.1}
\left[\ \boldsymbol{L},\ \boldsymbol{e}^{j}\cdot\boldsymbol{\omega}\ \right] = -\,i\hbar\ \boldsymbol{m}^{(j)}\left(\boldsymbol{\omega}\right)\,.
\end{equation}
The property~\eqref{3.39} and the commutation relation~\eqref{3.35.1} imply that the canonical commutation relations for the quantum description of a rotation are given by,
\begin{equation}\label{3.36.1}
\left[\ \boldsymbol{n}_{(j)}\left(\boldsymbol{\omega}\right)\cdot\boldsymbol{L},\ \boldsymbol{e}^{k}\cdot{\boldsymbol{\omega}}\ \right] = -\,i\hbar\,\left(\boldsymbol{e}_{j}\cdot\boldsymbol{e}^{k}\right)\,.
\end{equation}

In order to avoid any confusion, we would like to emphasize that the canonical commutation relations~\eqref{3.36.1} involve the orientation operator rather than the phase operator. For a one-dimensional harmonic oscillator, it is well-known that the latter is not self-adjoint since it is defined in terms of the position and momentum operators obeying canonical commutation relations~\cite{Carruthers:1968,Alimov:1979}. On the contrary, since the orientation operator $\boldsymbol{\omega}$ is a real function of the position operators, it is a self-adjoint operator.

\section{Heisenberg inequalities}

The vibrational canonical commutation relation~\eqref{3.32.1} implies the existence of vibrational Heisenberg inequalities, i.e.
\begin{equation}\label{3.37}
\Delta\,P_{\alpha}\,\Delta\,Q^{\beta} \geq \displaystyle{\frac{\hbar}{2}}\,\delta_{\alpha}^{\beta}\,.
\end{equation}
Similarly, the electronic canonical commutation relation~\eqref{3.32.1} implies the existence of electronic Heisenberg inequalities, i.e.
\begin{equation}\label{3.40}
\Delta\left(\boldsymbol{e}_{j}\cdot\boldsymbol{p}_{(\nu)}\right)\,\Delta\left(\boldsymbol{e}^{k}\cdot\boldsymbol{q}_{(\nu^{\prime})}\right) \geq \displaystyle{\frac{\hbar}{2}}\,\delta_{\nu\nu^{\prime}}\,\left(\boldsymbol{e}_{j}\cdot\boldsymbol{e}^{k}\right)\,.
\end{equation} 
Finally, the rotational canonical commutation relation~\eqref{3.36.1} implies the existence of rotational Heisenberg inequalities, i.e.
\begin{equation}\label{3.41}
\Delta\left(\boldsymbol{n}_{(j)}\left(\boldsymbol{\omega}\right)\cdot\boldsymbol{L}\right)\,\Delta\left(\boldsymbol{e}^{k}\cdot\boldsymbol{\omega}\right) \geq \displaystyle{\frac{\hbar}{2}}\,\left(\boldsymbol{e}_{j}\cdot\boldsymbol{e}^{k}\right)\,.
\end{equation} 
In order to compute explicitly the dispersions in the rotational Heisenberg inequalities~\eqref{3.41}, cyclic boundary conditions have to be taken carefully into account~\cite{Holevo:2011}.

The orbital angular momentum $\boldsymbol{L}$ and the corresponding dispersion $\Delta\left(\boldsymbol{n}_{(j)}\left(\boldsymbol{\omega}\right)\cdot\boldsymbol{L}\right)$ can be measured using a circularly polarised light beam scattered by the spinning molecules~\cite{Allen:1992} based on a recent technique involving a rotational Doppler shift~\cite{Lavery:2013}. The shift in frequency of circularly polarized light due to the scattering is proportional the angular velocity of the molecule $\boldsymbol{\Omega}$ and to the sum of the orbital angular momentum of the molecule $\boldsymbol{L}$ and the angular momentum of the light beam.

The molecular orientation axis $\boldsymbol{\omega}$ and the corresponding dispersion $\Delta\left(\boldsymbol{e}^{k}\cdot\boldsymbol{\omega}\right)$ can be measured using a strong laser pulse~\cite{Stapelfeldt:2003,Viftrup:2007}. The pulse induces an electric dipole along the direction of highest polarisability of the molecules. In order to minimise the electric dipolar energy, the dipoles align with the electric field of the laser pulse thus orienting the molecules.

\section{Conclusion} 

In order to obtain the Rotational Heisenberg Inequalities, we establish the quantum dynamics of an isolated molecular system where all the physical degrees of freedom are described by operators, including the rotational degrees of freedom that are defined by the Lie algebra of the rotation group.

Since there exists no rest frame in a quantum description of a molecular system, we used algebraic relations between the position and momentum observables associated to the nuclei and electrons in order to determine the position and momentum ``rest'' observables defined with respect to the rotating molecule. Recasting the ``rest'' observables in terms of internal observables accounting for the vibrational rotational and electronic degrees of freedom leads to one canonical commutation relation for each degree of freedom. These commutation relations yield vibrational electronic and rotational Heisenberg inequalities.

The Rotational Heisenberg Inequalities~\eqref{3.41} are the product of the molecular orbital angular momentum dispersion and the molecular orientation dispersion.

\bibliography{references}
\bibliographystyle{eplbib}

\end{document}